\documentclass[]{mn2e}

\usepackage{amssymb}
\usepackage{graphicx}
\usepackage{amsmath}
\usepackage{graphics}

\usepackage{color}

\title[Far infrared properties of graphite]{Some optical properties of graphite from IR to millimetric wavelengths}
\author[R. J. Papoular and R. Papoular]{Robert J. Papoular$^{1}$\thanks{E-mail:
Robert.Papoular@cea.fr} and 
Renaud Papoular$^{2}$\thanks{E-mail: papoular@wanadoo.fr}\\
$^{1}$IRAMIS, Laboratoire Leon Brillouin, CEA Saclay, 91191 Gif-s-Yvette, France\\
$^{2}$Service d'Astrophysique and Service de Chimie Moleculaire,
 CEA Saclay, 91191 Gif-s-Yvette, France}

\begin{document}

   \maketitle
\label{firstpage}

\begin{abstract}

Far infrared(FIR) data on the optical properties of graphite are presently lacking. An important step towards filling this gap was taken by Kuzmenko et al. \cite{kuz} who measured, on HOPG (Highly Oriented Pyrolitic Graphite) {\bf at normal incidence} and from 10 to 300 K, the in-plane dielectric functions from 0.3 to 200 $\mu$m, and the reflectance between 0.3 and about 300 $\mu$m. We show here how, using recent developments of the electron theory of graphene, extended to graphite, it is possible to properly extrapolate the data farther even than 1000 $\mu$m, in effect all the way to DC (Direct Current). The plasma frequency as well as the scattering rate of free electrons are shown  to both decrease with $T$, but level off near 0 K, in agreement with theory. Along the way, we noticed significant discrepancies with the well-known and often used derivation of Philipp \cite{phi77} at room temperature, and also with previous data on temperature dependence and absorbance of graphitic material samples in different physical forms. Possible reasons for these discrepancies are discussed. Finally, the absorption efficiency of small graphitic spheres is deduced for the spectral range from 0.3 to 10000 $\mu$m. This may contribute to the discussion on model dust candidates for recently observed astronomical far infrared emissions.

\end{abstract}

\begin{keywords}
astrochemistry---dust---FIR
\end{keywords}



\section{Introduction}

 The FIR (Far Infra Red) {\it Planck} astronomical observatory was launched in 2009 and delivered, in 2011, a treasure trove of FIR Galactic emission data (Tauber et al. 2010, Planck collaboration 2011). It has observed the sky in a wide frequency range, from the cosmic microwave band to the FIR with high sensitivity and angular resolution. In particular, the bolometers of its High Frequency Instrument, cooled to 0.1 K, cover the 100, 143, 217, 353, 535 and 857 GHz bands. They delivered exquisitely detailed and accurate measures of electro-magnetic emission from the Diffuse Galactic Interstellar Medium (DGISM) at high latitudes, which allowed the probing of the full Spectral Energy Distribution (SED) of the thermal emission of large dust grains that make up a sizable fraction of the InterStellar (IS) dust mass. These results complement and support previous space missions which contributed considerably to the same field, such as the Infra Red Astronomical Satellite (IRAS) and the Cosmic Background Explorer (COBE/FIRAS). Analyses of these measurements were made by Boulanger et al. \cite{bou96}, Abergel et al. \cite{abe11} and Compi{\`e}gne at al. \cite{com11}, who established that the average thermal spectrum of thin, quiescent clouds in the local DGISM could be fitted by a gray-body spectrum at a temperature in the range 17-19 K. This spectrum, therefore, essentially covers the range from 100 to 1000 $\mu$m. 

 The search for possible dust carriers of IS emission usually centers on graphite, or ``astronomical graphite and amorphous silicate''(Draine and Lee 1984). An example of more recent models of the latter is that of Compi{\`e}gne at al. \cite{com11}. This model includes bare silicate grains, PAHs and ``amorphous carbon dust'' (see their Table 2). For the latter, the authors used refractive indexes derived by Zubko et al. \cite{zub96} for a BE sample. This term designates a powder obtained by striking an electric arc from graphitic electrodes in an atmosphere of benzene or hydrogen, and was coined by Koike et al. \cite{koi80}, who qualified the material as amorphous carbon. The latter appellation  may be slightly confusing because, as a matter of fact, the grains making up the deposited powder were shown to consist of \emph{randomly} dispersed \emph{crystallites}, which electron microscopy shows are \emph{ordered} like graphite (see Koike et al. 1994), and not microscopically amorphous (like, for instance, HAC, hydrogenated carbon). Unfortunately, subsequent authors who used the Zubko model sometimes also insisted that they rejected graphite in favor of HAC (see Compi{\`e}gne et al. 2011, p. 4). 

This might give the impression that graphite \emph{per se} is no longer of use to astrophysics. Witness to the opposite is the wide use of Polycyclic Aromatic Hydrocarbon (PAH), when it is considered that very small particles of graphite are essentially dehydrogenated PAHs (Zubko et al. 2004). Even the latest model of Draine and Li \cite{dra07} includes graphite with properties from Draine and Lee \cite{dra84}. Besides, we ourselves recently showed the relevance of small imperfect/impure graphitic grains (carrying a minority of the available carbon atoms) to the modeling of the 2175 \AA \ features and its variations (Papoular et al. 2013). Thus, a full assessment of the role of graphite in astrophysics still requires some discussion.

The development of efficient techniques for the production of pure graphene layers (Geim and Novoselov 2007) and the subsequent demonstration of their exotic electrical behavior (see Castro-Neto et al. 2009) have spurred, in the technology community, a renewed interest in graphite, which is made of stacked graphene layers according to a definite geometrical rule (Bernal ABA stacking). This led Kuzmenko et al. \cite{kuz} to measure the reflectance of graphite at normal incidence, from about 1.8 to 300 $\mu$m, and at 30 temperatures between 10 K and ambient. They also used ellipsometry to complement these measurements down to 0.3 $\mu$m. During the past decade, great strides were made in the theoretical understanding of electron transport in graphene and graphite (see, for instance, Castro-Neto et al. 2009, Falkovsky and Varlamov 2007, Katsnelson and collegues: Yuan et al. 2011a,b and included bibliography). This is all of great help for our present purposes.

Now, of course, several measurements of graphite electrical properties were conducted since Hoyle and Wickramasinghe proposed this material as a major component of Interstellar (IS) dust (see Wickramasinghe 1967). These are technically difficult measurements, especially for the edge-on orientation ($\vec c \| \vec E$). Even face-on (in-plane) measurements have proved so challenging, especially in the FIR, that the results obtained by different authors rarely converge. Here, we compare some of these in an attempt to try and understand the causes of such discrepancies. We pay particular attention to  Philipp's work \cite{phi77}, which has been the basis of most astronomical uses of graphite as a model grain material.

The paper is organized as follows. Section 2 reports the measurements by Kuzmenko et al. \cite{kuz} and their interpretation in the light of recent theories of graphene, with particular attention to temperature effects. Section 3 describes the original procedure used here to obtain a physically sensible continuation of those measurements into the FIR, beyond 10$^{4}\,\mu$m . Section 4 illustrates the essential optical properties delivered by our treatment: dielectric functions, refractive indexes, and complex reflectance for 4 typical temperatures. Section 5 compares our absorption efficiency (Q/a) for graphite at 300 K with those of previous authors, identifies the discrepancies between different published results and discusses their possible origins. Section 6 finally briefly discusses the relevance of graphite to the modeling of FIR and millimetric IS emission, as compared to other models.

\section{The measurements of Kuzmenko et al. (2008)}

\begin{figure}
\resizebox{\hsize}{!}{\includegraphics{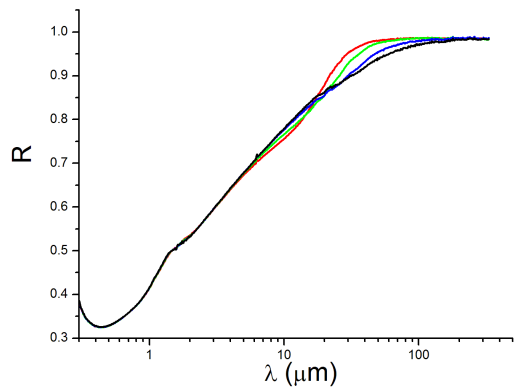}}
\caption[]{Normal incidence reflectance on HOPG measured by Kuzmenko et al. \cite{kuz} at 300 K (red, upper curve), 200 K (green), 100 K (blue) and 10 K (black, lower curve). The temperature effects are limited, essentially, to the range beyond 5 $\mu$m; in this range reflectance increases with temperature by less than 10 \%.}
\label{Fig:Rkuz}
\end{figure}

\begin{figure}
\resizebox{\hsize}{!}{\includegraphics{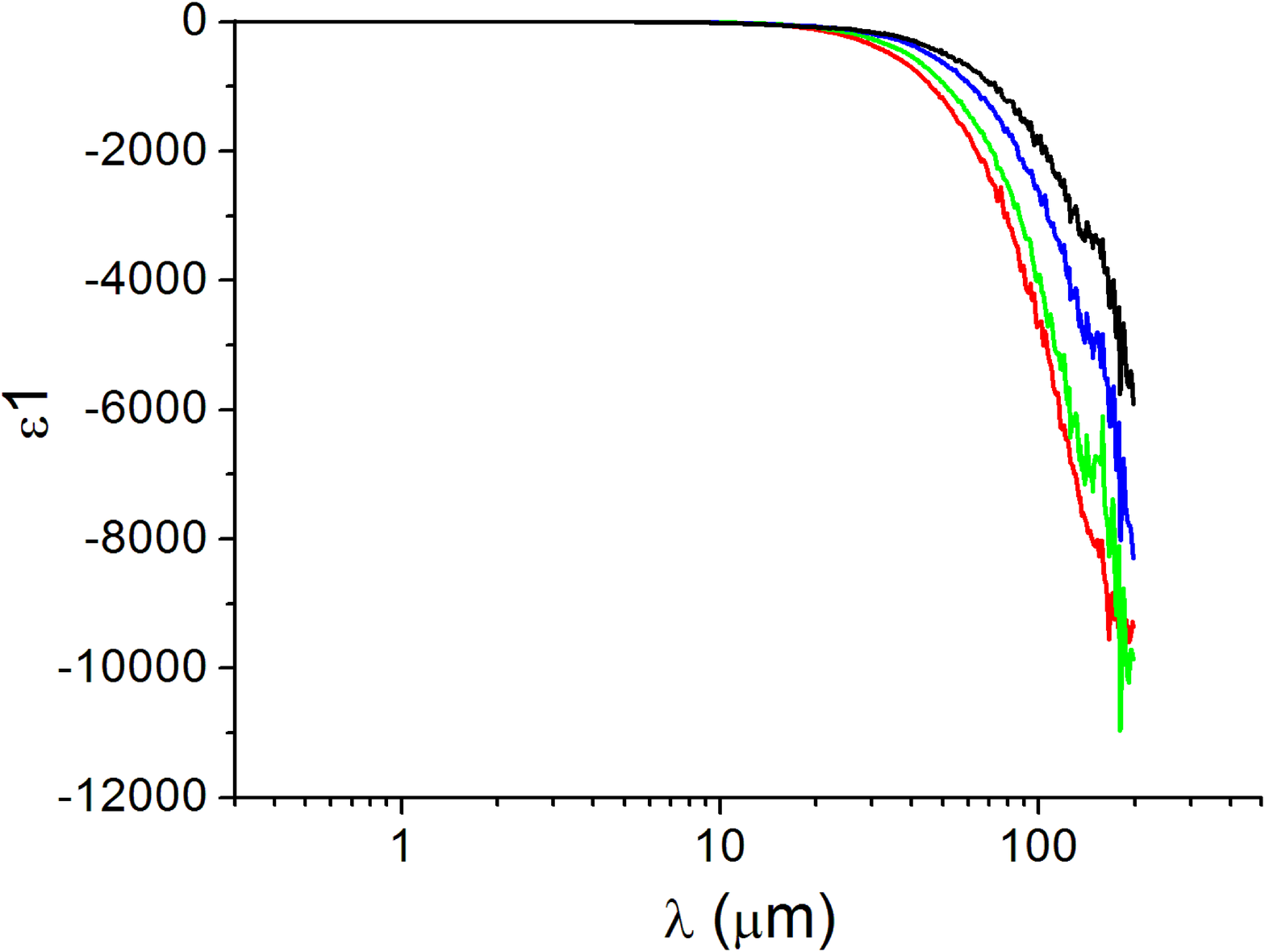}}
\caption[]{Real part of the dielectric function of HOPG measured by Kuzmenko et al. \cite{kuz} at normal incidence, at 300 K (red, lower curve), 200 K (green), 100 K (blue) and 10 K (black, upper curve). The temperature effects are limited, essentially, to the range beyond 5 $\mu$m.}
\label{Fig:e1kuz}
\end{figure}

\begin{figure}
\resizebox{\hsize}{!}{\includegraphics{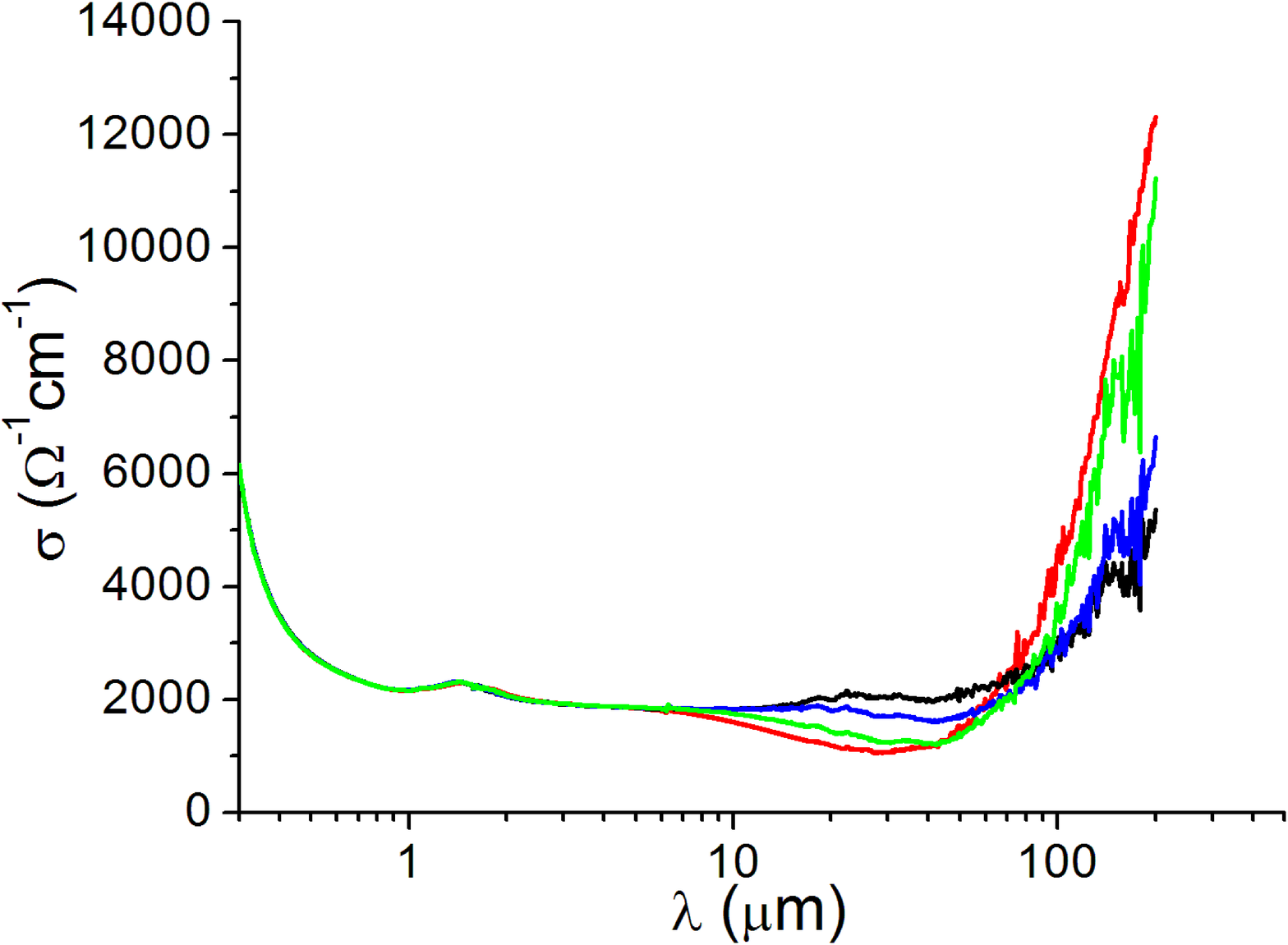}}
\caption[]{Real part of the conductivity of HOPG measured by Kuzmenko et al. \cite{kuz} 
at normal incidence, at 300 K (red, upper curve), 200 K (green), 100 K (blue) and 10 K (black, lower curve). The temperature effects are limited, essentially, to the range beyond 5 $\mu$m.}
\label{Fig:sigmakuz}
\end{figure}

An essential basis of our work is the data measured or deduced by Kuzmenko et al. \cite{kuz}, in the form of files kindly provided by A. Kuzmenko, for the normal incidence reflectance, $R$, in-plane conductivity, $\sigma$ and real part of the dielectric function, $\epsilon_{1}$. The graphical representation of these is given in Fig. \ref{Fig:Rkuz}, \ref{Fig:e1kuz} and \ref{Fig:sigmakuz}, where, for clarity, only 4 typical temperatures, 10, 100, 200 and 300 K, were retained from the 30 that were given.

The imaginary part of the dielectric function, $\epsilon_{2}$, is derived from the conductivity by (see Bohren and Huffman 1983)

\begin{equation}
\displaystyle
\epsilon_{2}=6 \,10^{-3} \,\sigma(\Omega^{-1}\mathrm{cm}^{-1})\,\lambda(\mu\mathrm{m}).
\end{equation}
From this, one can derive the refraction indexes $n,k$, the reflectance phase $\theta$, the absorbance $\alpha=4\pi k/\lambda$, and the absorption/extinction efficiency of small grains of radius $a$, $Q/a$ (see Bohren and Huffman 1983).

\begin{figure}
\resizebox{\hsize}{!}{\includegraphics{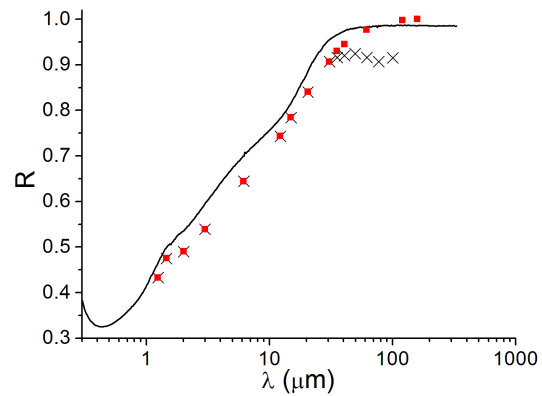}}
\caption[]{\it Black line\rm: Normal reflectance on graphite measured at room temperature by Kuzmenko et al. \cite{kuz}. \it Red dots \rm: the same by Taft and Philipp \cite{taf65}; \it Crosses \rm: the same by Philipp \cite{phi77}, who considered two possible extrapolations beyond about 40 $\mu$m: crosses or red squares (see Philipp's Fig. 1); he finally retained the crosses.}
\label{Fig:Rphil77}
\end{figure}

Next, Fig. \ref{Fig:Rphil77} pinpoints one of the primary motivations of this work, to wit the differences between the reflectance of Kuzmenko et al. \cite{kuz} and those of Taft and Philipp \cite{taf65}, and Philipp \cite{phi77}, all at room temperature. Clearly, the dielectric functions beyond about 20 $\mu$m cannot agree. It must be remembered that Philipp's extrapolation beyond 40 $\mu$m was based not on his own measurements, but on previous measurements and analysis by Sato \cite{sat}. The leveling off at $R\sim0.9$, followed by Philipp's Drude continuation to longer wavelengths, is the cause of a dramatic dip/peak in the dielectric functions, which reverberated strongly in Draine and Lee's ``astronomical graphite" \cite{dra84}. The  current graphite measurements and theories predict quite a different behavior beyond 10 $\mu$m, as shown below.

\section{The continuation into the far infrared}
Absent experimental results beyond 200 $\mu$m, understanding the physics of electron transport in graphite is a prerequisite to a sensible continuation of the data of Kuzmenko et al. \cite{kuz} into the far infrared. This is made far easier by the recent elucidation of the properties of graphene, as graphite is obtained by stacking single graphene layers on top of each other according to a definite scheme (Bernal's ABA). 

At 0 K, the conductivity of \emph{graphene} beyond the $\pi$ resonance, through to the FIR, is constant and equal to $\sigma_{0,2D}=\frac{e^{2}}{4\hbar}=6.08\,10^{-5}\,\mho$ (see Kuzmenko 2008), where $e$ is the electron charge. Remarkably, this expression does not depend on any structural parameter. It is a result of the conical shape (as opposed to the usual parabolic shape) of the valence and conductivity electronic bands of graphene in reciprocal space, near the Dirac point K of its Brillouin zone, where they meet at the common apex of the two opposite cones. 

In \emph{graphite} at 0 K, in first approximation, this translates into a constant conductivity

\begin{eqnarray}
\displaystyle
\sigma_{0,3D } &=& \sigma_{0,2D}/d  \nonumber   \\
                         &=& 1810 \, \mho \mathrm{cm^{-1}} =1.63 \,10^{15}\mathrm{s^{-1}}=1.07 \mathrm{eV}.
\end{eqnarray}

where $d=3.36\,{\rm \AA}{\ }$ is the interlayer spacing. This expression, too, does not depend on any structural parameter other than $d$. However, electron hopping between the constitutive monolayers of graphite introduces subtle changes in the \mbox{geometry} of electronic bands in reciprocal space, giving rise to a ``self-doped material" (see Kuzmenko 2008). As a consequence, graphite at 0 K displays only roughly constant conductivity and then only in a limited spectral range (1 to 100 $\mu$m, designated below as the ``intermediate plateau"). In this range, Figure \ref{Fig:sigmakuz}, for T=10 K, indeed displays a segment with nearly constant ordinate, except for two weak bumps discussed in more details by Kuzmenko et al. \cite{kuz}, but overlooked here. Obviously, the rising segment on the left is the red wing of the $\pi$ resonance which peaks near 4.1 eV or about 0.3 $\mu$m.

Now, when the temperature rises above 0 K, both in graphene and graphite, more and more electrons are freed from the valence band into the conduction band, leaving behind an equal number of holes. The resulting increase in conductivity can be described analytically by a Drude function (see Wallace 1947, Pedersen 2003, Falkovsky and Varlamov 2007), which rises to a  ``FIR plateau", extending to DC, whose height increases with temperature. Although this part of the FIR conductivity is not visible in Fig. \ref{Fig:sigmakuz}, it is apparent that the measurements reach a long way towards it, and this will be of great help for our present extrapolation purposes. Note that the FIR plateau is due to \emph{intra}band conductance, while the intermediate plateau, as well as the resonances, is due to \emph{inter}band conductance.

As the number of electrons {\bf per cell} of the material is finite and constant (f-sum rule), the increased optical weight in the FIR must be exactly compensated by a decrease in the adjacent part of the conductivity plateau, as evidenced by dips below the average plateau in the same figure. An adequate continuation into the FIR must account for both the increase and decrease of the conductivity (in different but adjacent parts of the IR spectrum), as the temperature varies.

 While the conductivity of \emph{graphene} at 0 K is theoretically constant throughout the IR,  another consequence of inter-layer hopping in \emph{graphite} is that the FIR conductivity plateau (beyond 1000 $\mu$m) is higher than the intermediate plateau (roughly 1 to 100 $\mu$m), even at 0 K, as suggested by the curve for T=10 K in Fig. \ref{Fig:sigmakuz}.

The above indicates that a first step, in extrapolating the experimental results into the FIR, is to use a Drude function, as Philipp \cite{phi77} did before:
\begin{equation}
\displaystyle
\sigma(\omega)=\epsilon_{0}\frac{\omega_{p}^{2}\gamma}{\omega^{2}+\gamma^{2}}\,,
\end{equation}
where $\omega$ is the frequency, $\epsilon_{0}$ the dielectric constant of vacuum, $\omega_{p}$ the plasma frequency and $\gamma$ the dispersion constant or line width, physically associated with electron scattering. While $\omega_{p}$ obviously increases with temperature $T$, $\gamma$ may be a function of $T$ but should remain in the vicinity of 6 mev ($\lambda\sim200\,\mu$m), where the curves of Fig. \ref{Fig:sigmakuz} all quickly rise.

 However, one cannot stay content with this. In order to account for the central plateau and its deformation as $T$ increases, we therefore added to the conductivity a ``negative" modified Drude function 
\begin{equation}
\displaystyle
\sigma_{1}(\omega)=-\frac{C\sigma_{0}\gamma_{1}^{2}}{\omega^{2}+\gamma_{1}^{2}},
\end{equation}
where $\sigma_0$ is identical with $\sigma_{0,3D}$ of eq. 2 and   $\gamma_{1}$ may be a function of $T$. This choice of function is empirical and devoid of physical meaning, but guided by the need for a function whose conjugate by the Kramers-Kronig transform is known. Similarly, for the left part of Fig. \ref{Fig:sigmakuz}, where the intermediate plateau must fade away into the $\pi$ resonance we represented it by the sum of the characteristic Lorentzian of this resonance,
\begin{equation}
\displaystyle
\sigma_{\pi}(\omega)=\epsilon_{0}\frac{\omega_{p\pi}^{2}\gamma_{\pi}\omega^{2}}{(\omega_{0\pi}^{2}-\omega^{2})^{2}+\gamma_{\pi}^{2}\omega^{2}},
\end{equation}
 and a ``positive" modified Drude function 

\begin{equation}
\displaystyle
\sigma_{2}(\omega)=\frac{C\sigma_{0}\gamma_{2}^{2}}{\omega^{2}+\gamma_{2}^{2}},
\end{equation}

\noindent
where $\gamma_{2}$ need not vary with $T$ as all curves in Fig. \ref{Fig:sigmakuz} merge together, independant of $T$. The adjustable dimensionless parameter $C$, common to both modified Drude functions, defines the height of the plateau and its value should theoretically remain close to 1. 

This choice of modeling functions allows one to promptly deduce the real and imaginary parts of the dielectric functions and, hence, all optical quantities of interest. It is then possible to tailor the parameters defining the 4 functions, so as to best fit the experimental results. The satisfactory fits obtained this way, as shown below, give confidence in this representation, at least for practical purposes, which has three benefits: first, one can dispense with the heavy burden, and attendant uncertainties at both ends of the spectral window considered, of applying the Kramers-Kronig relations to infer, for instance, the phase of the reflectivity from the measured reflectance, $R$, so as to deduce the optical functions; second, all 3 given sets of experimental data are simultaneously used to fit the measurements; three, it is possible to define analytically the dielectric functions, once the temperature dependences of the parameters have been found (empirically or theoretically).

\section{Results}

As explained above, all experimental optical properties of interest here can be deduced directly from any pair among the 3 quantities $R$, $\epsilon_{1}$ and $\sigma$ for which
 Kuzmenko et al. \cite{kuz} gave experimental data up to 200 $\mu$m; they are therefore all labeled below as ``measured". For a given temperature, a fit of our model to any of the latter two delivers all other optical quantities. In effect, the fits were optimized by simultaneous comparison of experimental and model spectra of $\epsilon_{1}, \epsilon_{2}, \sigma/\sigma_{0}, R, \theta, n$ and $k$ in the hope of compensating possible measurement uncertainties. The fits were then perfected by optimizing  the parameters using non-linear least-squares minimization based on the Davidon-Fletcher-Powell (DFP) method \cite{fp}.

The following figures illustrate the spectra delivered by the model (dashes) and superimposed upon the corresponding measured spectra (dots), using black, blue, green and red colors for 10, 100, 200 and 300 K, respectively. 

Table 1 collects the fit parameters for the 4 temperatures. In all cases, the best fit to the plateau was obtained by adjusting the value of C in the expressions for the ``modified" Drude functions. The parameters of the $\pi$ Lorentzian are very close to those adopted by Papoular and Papoular \cite{pap09} to fit the observed UV feature at 2175 \AA. {The fit parameters for all 30 temperatures measured by Kuzmenko et al. are given in the Appendix.}

 \begin{figure}
\resizebox{\hsize}{!}{\includegraphics{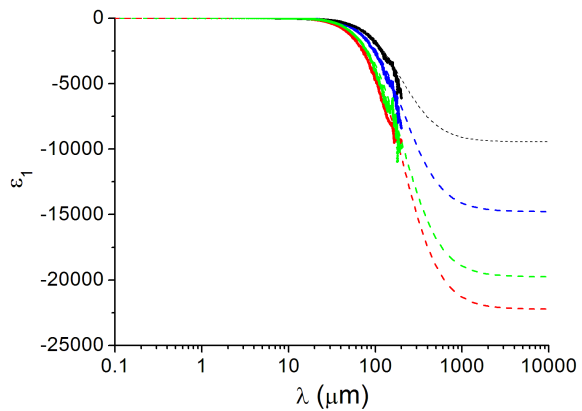}}
\caption[]{Real part of the dielectric function of HOPG measured by Kuzmenko et al. \cite{kuz} 
at normal incidence, at 300 K (red dots, lower curve)), 200 K (green dots), 100 K (blue dots) and 10 K (black dots, upper curve), superimposed upon the corresponding model spectra in dashed lines of corresponding colors.}
\label{Fig:e1}
\end{figure}

\begin{figure}
\resizebox{\hsize}{!}{\includegraphics{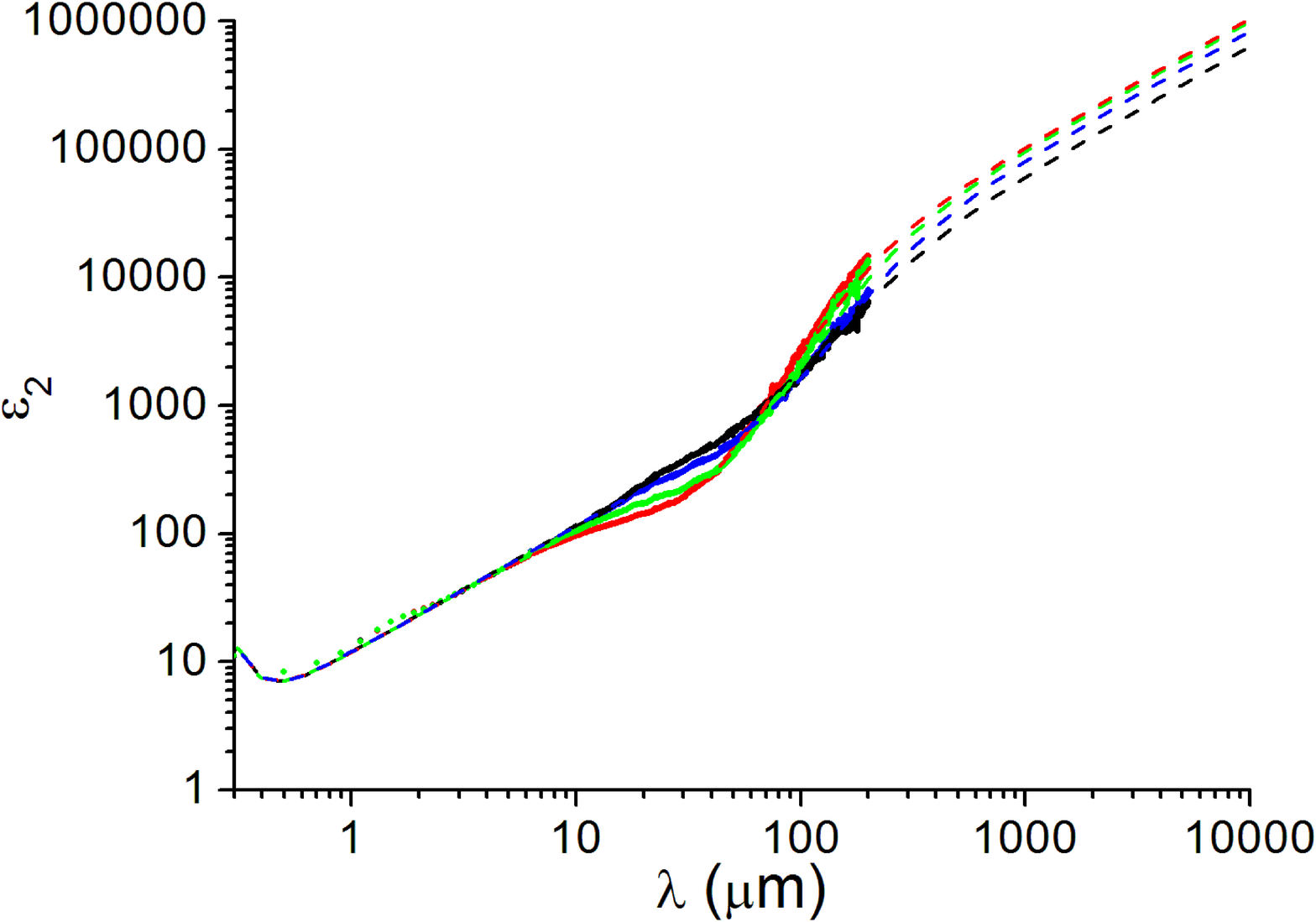}}
\caption[]{Imaginary part of the dielectric function of HOPG as measured by Kuzmenko et al. \cite{kuz} at normal incidence, at 300 K (red dots, upper curve), 200 K (green dots), 100 K (blue dots) and 10 K (black dots, lower curve), superimposed upon the corresponding model spectra in dashed lines of corresponding colors.}
\label{Fig:e2}
\end{figure}

\begin{figure}\resizebox{\hsize}{!}{\includegraphics{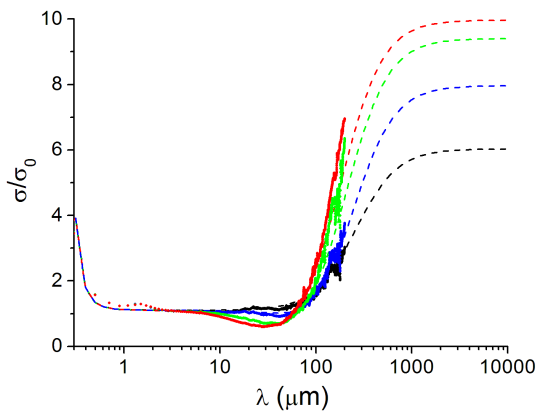}}
\caption[]{Real part of the conductivity of HOPG as measured by Kuzmenko et al. \cite{kuz} 
at normal incidence, at 300 K (red dots, upper curve), 200 K (green dots), 100 K (blue dots) and 10 K (black dots, lower curve), superimposed upon the corresponding model spectra in dashed lines of corresponding colors. Note: the curves here represent $\sigma/\sigma_{0}$, unlike in Fig. \ref{Fig:sigmakuz}.}
\label{Fig:sss0}
\end{figure}

\begin{figure}
\resizebox{\hsize}{!}{\includegraphics{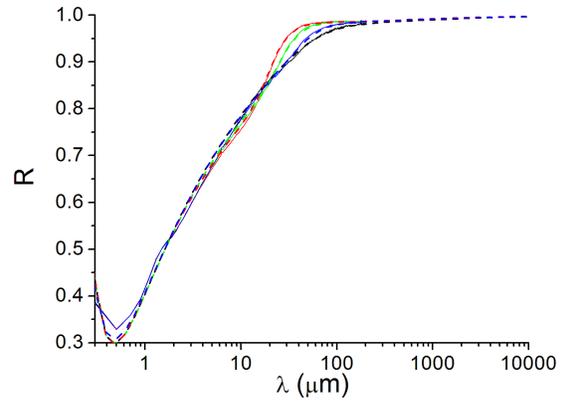}}
\caption[]{The reflectance of HOPG as measured by Kuzmenko et al. \cite{kuz} at normal incidence, at 300 K (red dots, upper curve), 200 K (green dots), 100 K (blue dots) and 10 K (black dots, lower curve), superimposed upon the corresponding model spectra in dashed lines of corresponding colors.}
\label{Fig:R}
\end{figure}

\begin{figure}
\resizebox{\hsize}{!}{\includegraphics{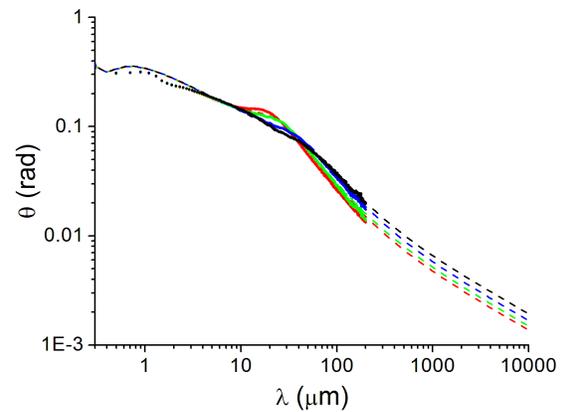}}
\caption[]{The phase, or argument, of the complex reflectance, $r$, whose modulus is the square root of the reflectance, $R$, of HOPG measured by Kuzmenko et al. \cite{kuz} 
at normal incidence, at 300 K (red dots, lower curve), 200 K (green dots), 100 K (blue dots) and 10 K (black dots, upper curve), superimposed upon the corresponding model spectra in dashed lines of corresponding colors.}
\label{Fig:ph}
\end{figure}

\begin{figure}
\resizebox{\hsize}{!}{\includegraphics{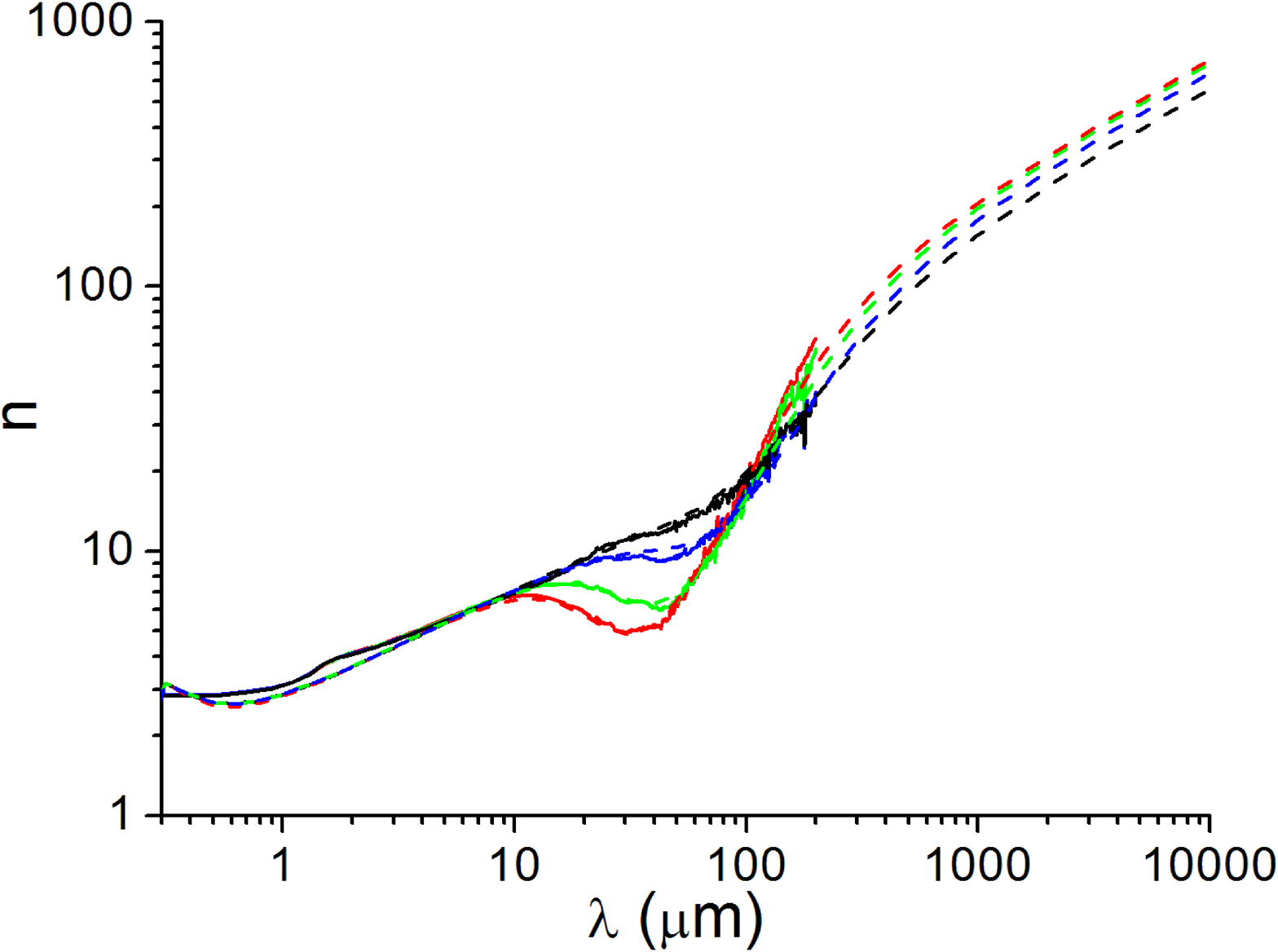}}
\caption[]{The real part of the refractive index of HOPG as measured by Kuzmenko et al. \cite{kuz} at normal incidence, at 300 K (red dots, upper curve), 200 K (green dots), 100 K (blue dots) and 10 K (black dots, lower curve), superimposed upon the corresponding model spectra in dashed lines of corresponding colors.}
\label{Fig:n}
\end{figure}

\begin{figure}
\resizebox{\hsize}{!}{\includegraphics{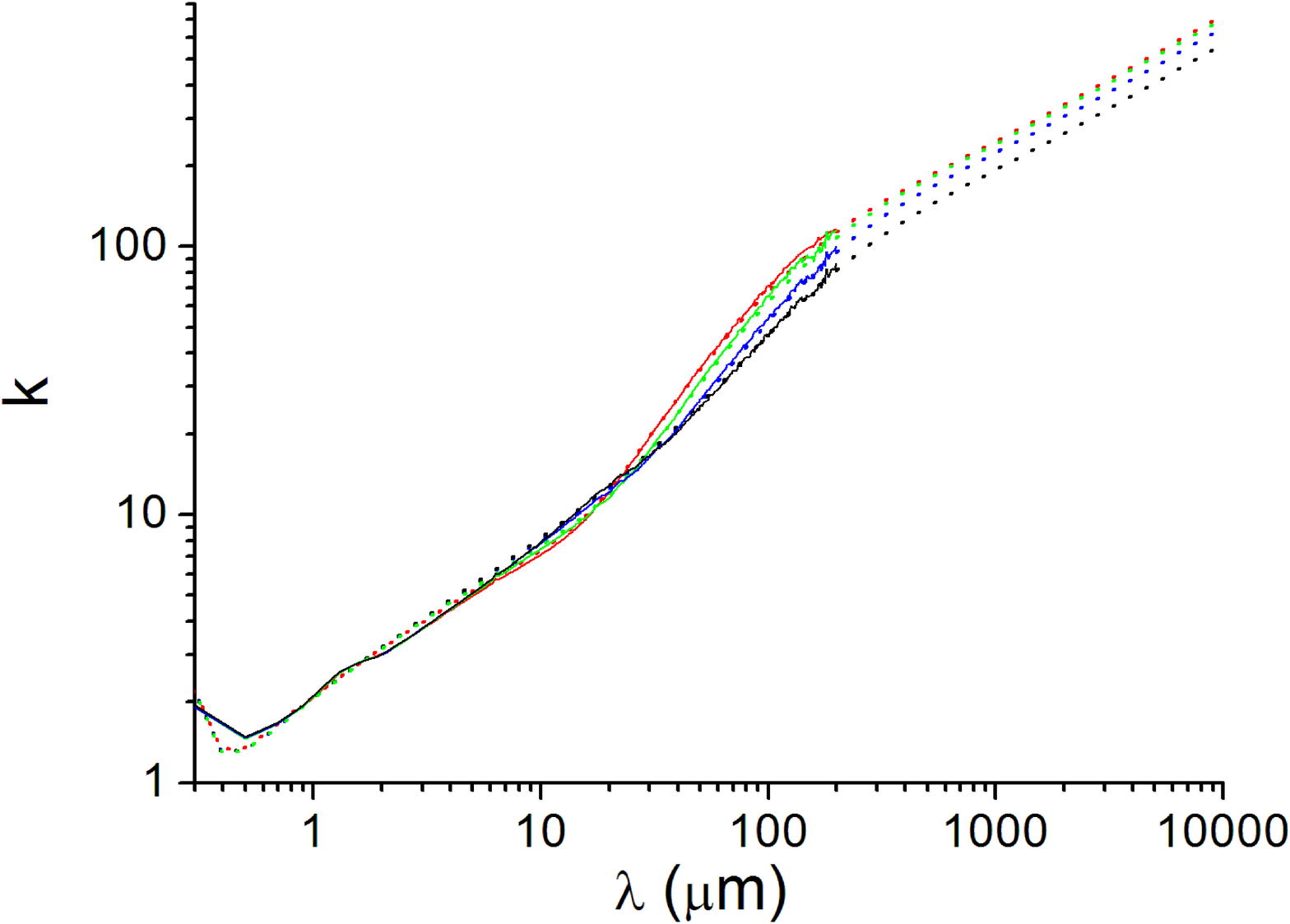}}
\caption[]{The imaginary part of the refractive index of HOPG as measured by Kuzmenko et al. \cite{kuz} at normal incidence, at 300 K (red dots, upper curve), 200 K (green dots), 100 K (blue dots) and 10 K (black dots, lower curve), superimposed upon the corresponding model spectra in dashed lines of corresponding colors.}
\label{Fig:k}
\end{figure}

\begin{table*}
\caption[]{Best fit parameters}
\begin{flushleft}
\begin{tabular}{lllllllll}
\hline
$T(K)$ & $\omega_{p}$ & $\gamma$ & $\gamma_{1}$ & $\gamma_{2}$ & $\omega_{0,\pi}$ & $\omega_{p,\pi}$ & $\gamma_{
\pi}$ & $W({\rm 12.6 \ meV})$\\
\hline
10 & 0.617 & 0.0048 & 0.0048 & 10 & 4.42 & 9 & 1.5 & 0.042\\
\hline
100 & 0.738 & 0.0052 & 0.0190 & 10 & 4.42 & 9 & 1.5 & 0.05\\ 
\hline
200 & 0.859 & 0.006 & 0.0427 & 10 & 4.42 & 9 & 1.5 & 0.064\\
\hline
300 & 0.944 & 0.0068 & 0.0627 & 10 & 4.42 & 9 & 1.5 & 0.073\\
\hline
W(12.6 meV)=$\int\sigma(\omega)d\omega/\sigma_{0,3D}$ is the total\\
optical weight between 0 and 12.6 mev ($\sim100 \mu$m).\\
All other values in eV.
\end{tabular}
\end{flushleft}
\end{table*}

 The fits generally appear to be adequate, except for $\sigma/\sigma_{0}$ at 200 and 300 K, and only between 150 and 200 $\mu$m, where the discrepancy reaches about $\sim20$ \%; naturally, this is also the case for the corresponding $\epsilon_{2}$ curves (Fig. \ref{Fig:e2}).
 
\begin{figure}
\resizebox{\hsize}{!}{\includegraphics{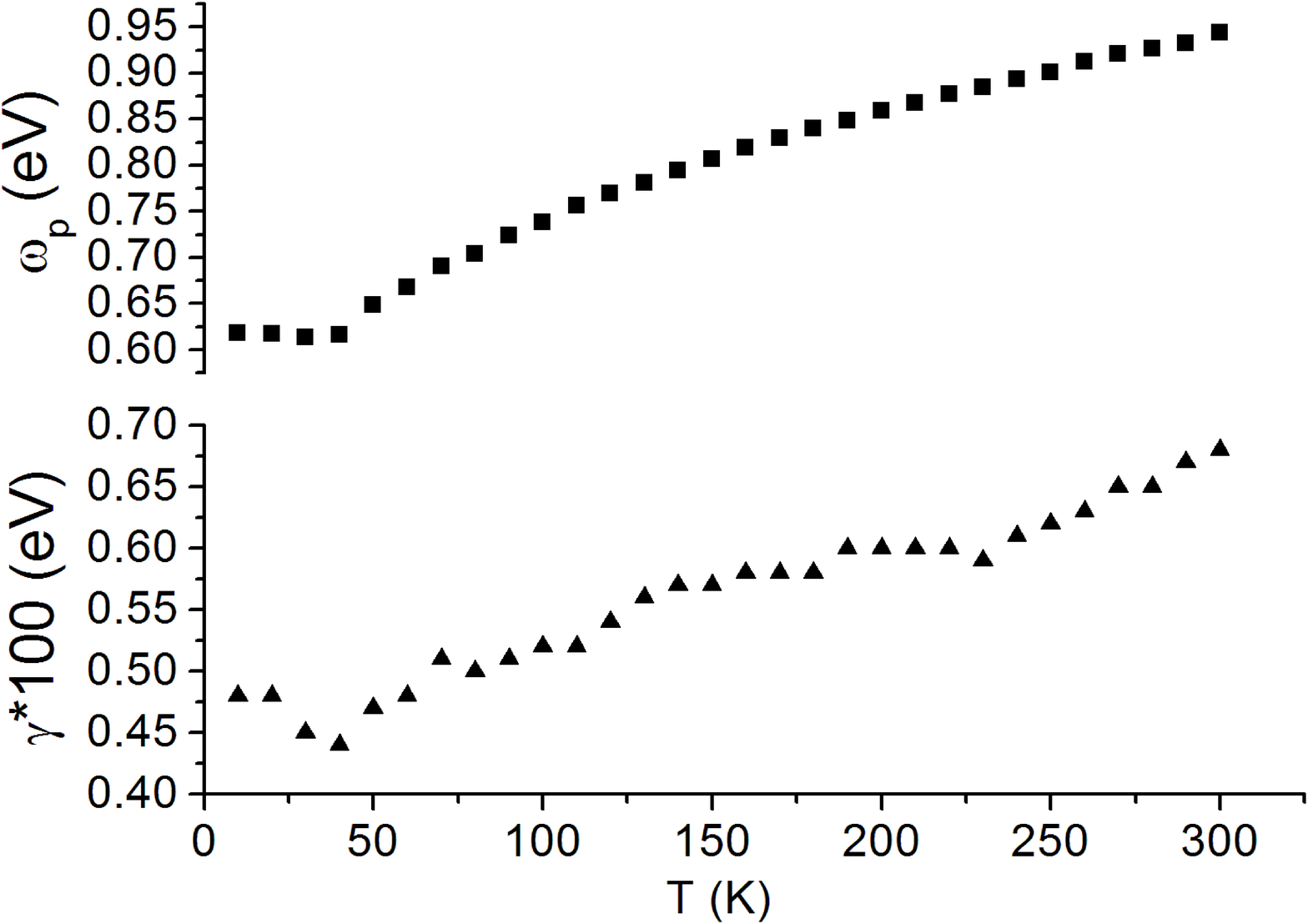}}
\caption[]{The plasma frequency, $\omega_{p}$ (squares) and the broadening, $\gamma$ (triangles), both in eV (for all 30 temperatures studied by Kuzmenko et al. \cite{kuz}). Their ratio remains nearly constant at $\sim140$.}
\label{Fig:wpga}
\end{figure}

 Beyond about 30 $\mu$m, both measured and model reflectances increase with temperature. This is where free electrons start contributing. This translates into an increase of the plasma frequency, $\omega_{p}$ (Fig. \ref{Fig:wpga}) which reflects the excitation of electrons above the Fermi level (see Gr\"uneis, 2008). The finite number of free electrons at very low temperatures is characteristic of semimetals like graphite, which is not the case of graphene (see Kuzmenko et al. 2008).

\begin{figure}
\resizebox{\hsize}{!}{\includegraphics{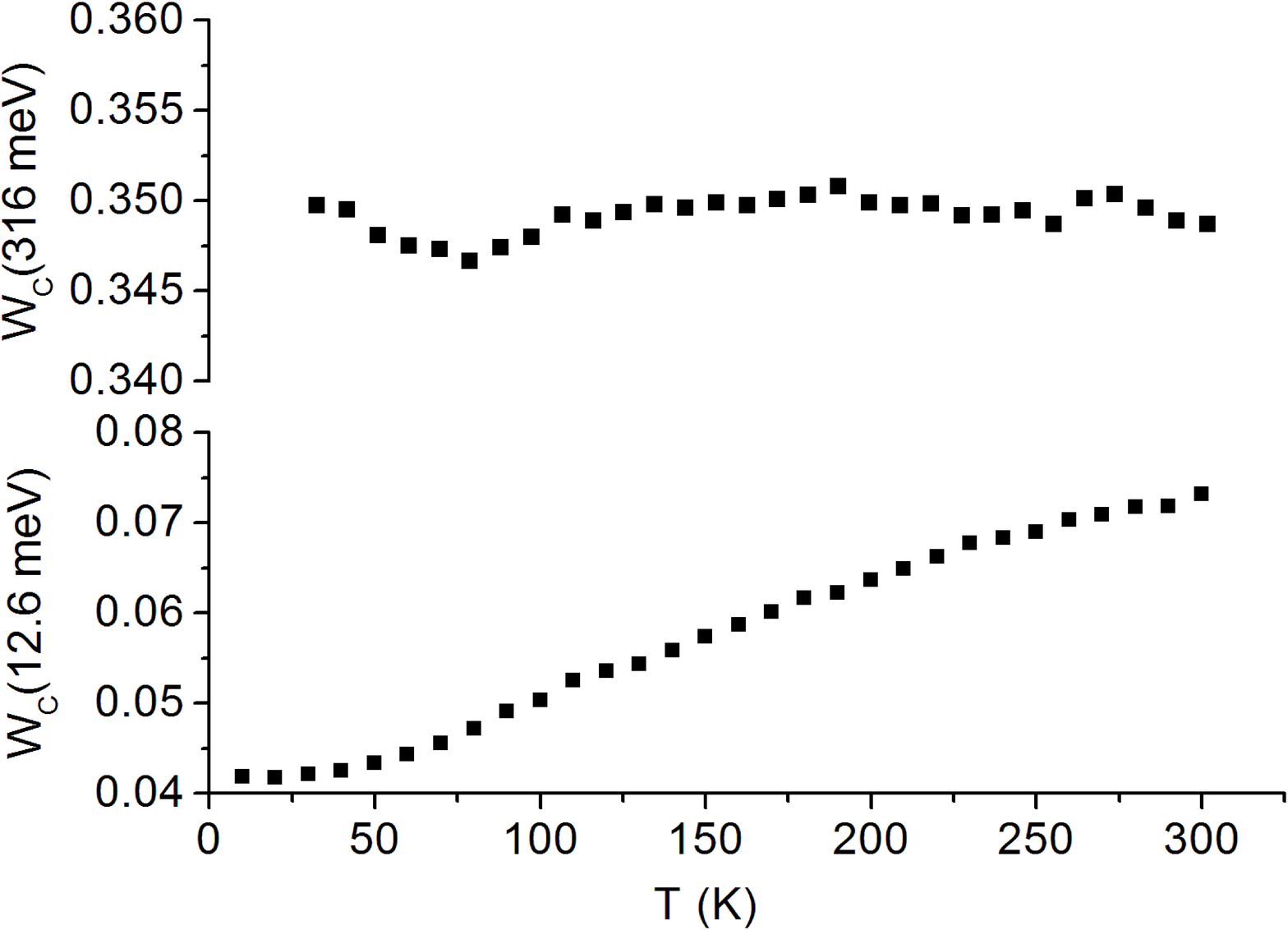}}
\caption[]{The optical weights W(316 \ meV) and W(12.6 \ meV). Note the near constancy of the former, indicating that the weight lost in the intermediate plateau (roughly 1 to 100 $\mu$m)
 of the conductivity,$\sigma$, is integrally transferred to the free electrons at longer wavelengths.}
\label{Fig:Wc}
\end{figure}

The broadening, $\gamma$, due to the free electrons being scattered, also increases, but only by 40 \% between 10 and 300 K. This is also the case for $\gamma_{1}$ as it should be, since it is tailored to account for the loss of available electrons in the valence band to the benefit of the plasmon. In this respect, our model is further validated by the fact that the optical weight from zero frequency to the middle of the plateau, W(0.316eV) (not included in Tab. 1), is indeed very nearly independent of temperature (Fig. \ref{Fig:Wc}) in accordance with the sum rule. 
 
 The slow variation of $\gamma$ with temperature, especially at very low temperatures,  indicates that the main cause of scattering of the free electrons is not impurities or phonons in their way. It might rather be collisions with the boundaries of the microcrystallites which constitute the building blocks of the HOPG sample. Here, the collision time, $1/\gamma$, is of the order of 10$^{-13}$ s and the free electron velocity at the Fermi level, about 10$^{8}$ cm.s$^{-1}$, so the mean free path is about 1000 \AA{\ }.

 Our values of W(0.316 eV) and W(12.6 meV) are only 20 \% higher than those of Kuzmenko et al. \cite{kuz}, Fig. \ref{Fig:Rphil77}: the inclusion, in W, of longer wavelengths (lower energies), attendant to our continuation into the FIR does not add much optical weight to the W's.

\begin{figure}
\resizebox{\hsize}{!}{\includegraphics{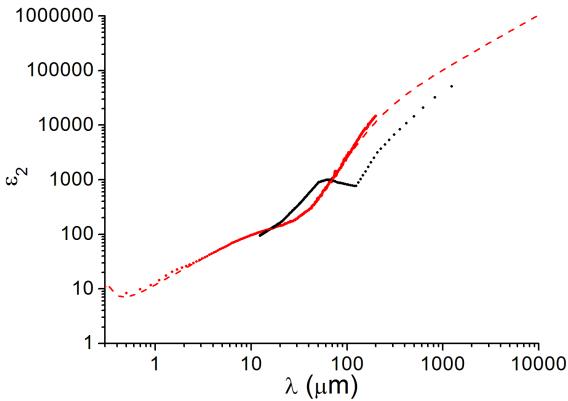}}
\caption[]{Imaginary part of the dielectric function of HOPG at 300K. \it Red line: \rm as measured by Kuzmenko et al. \cite{kuz} at normal incidence.\it Red dashes: \rm the corresponding model spectrum. \it Black dots: \rm adapted from Philipp \cite{phi77}.}
\label{Fig:e2phi77}
\end{figure}

Most importantly, but not unexpectedly, our dielectric functions differ considerably from Philipp \cite{phi77}: first, our $\epsilon_{1}$ lacks his narrow peak near 70 $\mu$m (18 meV);  second, the corresponding wiggle in $\epsilon_{2}$ is reversed with respect to the undulation that is physically due to the transfer of optical weight from bands to free electrons (our Fig. \ref{Fig:e2phi77}). Both accidents in Philipp's curves are, in fact, due to his choice of extrapolation of his reflectance curve. This will also reverberate into the behavior of $Q/a$, as shown in the next Section. Note that the reflectance previously measured by Taft and Philipp \cite{taf65}, as extrapolated into the FIR (Philipp 1977, Fig. 1) appears to increase continuously towards 1. Philipp \cite{phi77} subsequently modified that extrapolation, based on the reflectance measurements of Sato \cite{sat}, which first displayed this wiggle (see his Fig. 1).
 
These differences are, of course, linked with Philipp's preferred values of his Drude function:  $\omega_{p}=0.44\,$ eV, and $\tau=2\,10^{-13}\,$s (or $\hbar/\tau$=0.0032 eV) to be compared, respectively, with ours: 0.944 and 0.0068 eV.

\begin{figure}
\resizebox{\hsize}{!}{\includegraphics{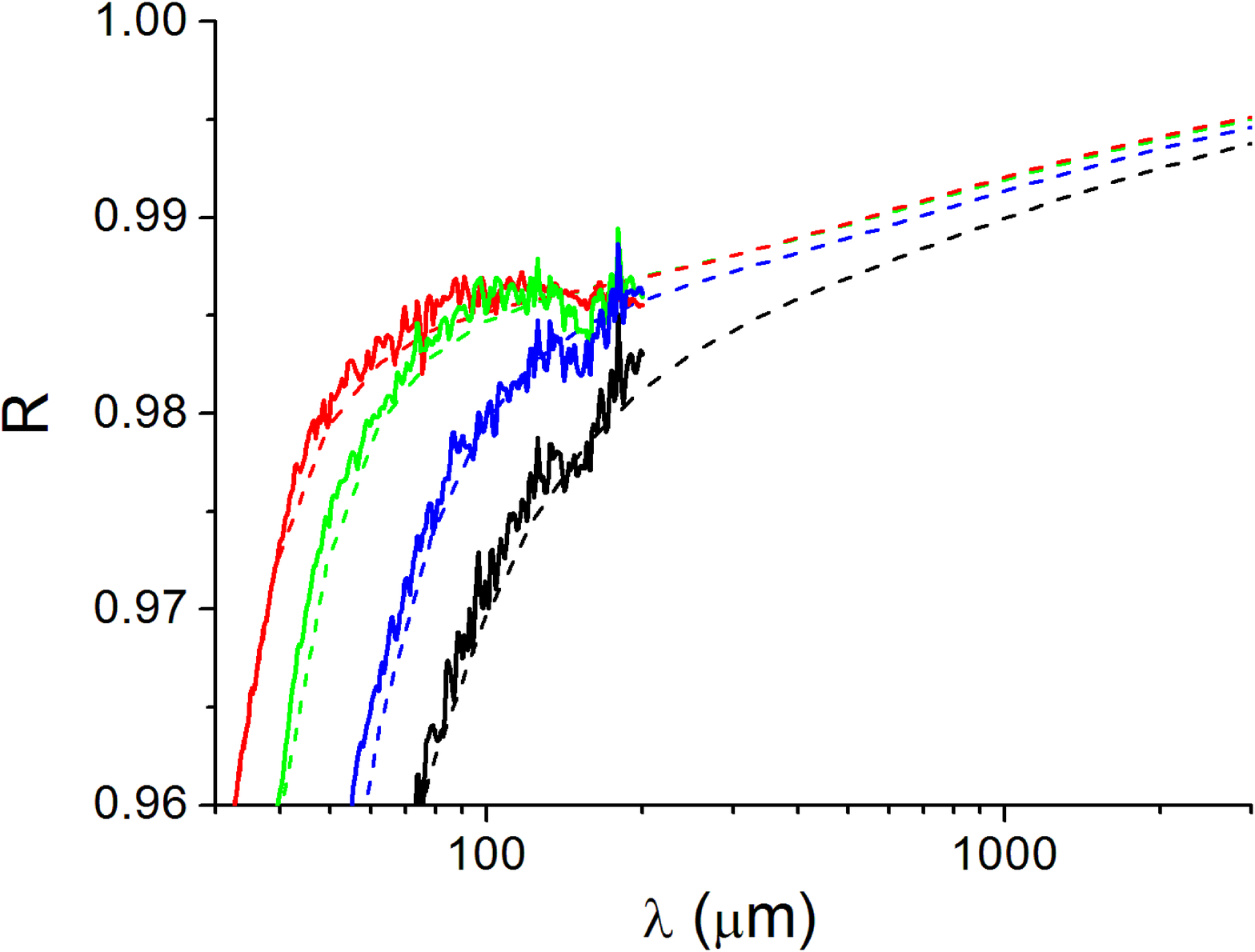}}
\caption[]{A zoom on the reflectance in the FIR.}
\label{Fig:Rzoom}
\end{figure}

It may be of interest to elaborate on the behavior of the reflectance in the FIR by studying the analytical properties of the Drude function $per se$. It shows that, for small values of $\omega_{p}$ and $\gamma$, R extrapolates smoothly and asymptotically to 1 as $\lambda$ increases, like the Taft and Philipp extrapolation \cite{phi77} in Fig. \ref{Fig:Rphil77} (red squares). As $\omega_{p}$ increases, a knee forms at shorter wavelengths; in between, R  levels off into a plateau-like segment starting near $\omega=\gamma$. At this point, it is found that
\begin{equation}
\displaystyle
R(\gamma)=1-1.82\frac{\gamma}{\omega_{p}}.
\end{equation}
Table 1 gives $\frac{\gamma}{\omega_{p}}\approx 0.0074 $ for the 4 values of T, so $R({\gamma})=0.9863$, in agreement with the behavior of Kuzmenko's experimental curves at their highest (Fig. \ref{Fig:Rzoom}).

By definition, the DC conductivity is the asymptotic value of $\sigma$: $\epsilon_{0}\omega_{p}^{2}/\gamma$. At 300 K, Fig. \ref{Fig:sss0} and Tab. 1 show this to be $10.02\times 1810=18135\,\mho\mathrm{cm^{-1}}$, or 1.63$\,10\,^{16}\,$s$^{-1}$. This is not significantly different than the value 2.5$\,10^{4}\,\mho\mathrm{cm^{-1}}$ as measured by Soule \cite{sou} or 2$\,10^{4}\,\mho\mathrm{cm^{-1}}$ as later reported by  Klein \cite{kle}.

However, we stress the different behaviors of the DC conductivity as the temperature changes, as inferred here from the measurements of Kuzmenko et al. \cite{kuz}, and as described much earlier by Soule \cite{sou}: the latter finds a smooth decrease of DC conductivity from 2$\,10^{5}$ at 20 K to about 2.5 $\,10^{4}\,\mho\mathrm{cm^{-1}}$, at room temperature (his Fig. 4). This decreasing trend was already predicted by Wallace \cite{wal} on the grounds that, due to electron scattering on phonons, the electron mean free path should decrease faster than $T^{-1}$, at variance with our findings.

 Several authors also found a positive temperature coefficient for the DC conductivity below room temperature: Noyes \cite{noy}, Buerschaper et al. \cite{bue}, Maltseva and Marmer \cite{mal}, Klein and Straub \cite{kle61}, Reynolds et al. \cite{rey}, Tyler et al. \cite{tyl}, Smith and Rasor \cite{smi}. The latter three groups, in particular, observed that this tends to occur in artificial graphites (like HOPG), with crystallites much smaller in size than natural graphite crystals. In that case, they argued, following a suggestion by Bowen \cite{bow},  electron scattering is dominated by collision with boundaries or simply lattice imperfections. 
The values of $\gamma$ we find in the present case suggest a crystallite size of order 1000 \AA \ $\perp\vec c$, which is the right order of magnitude for common artificial graphites. The temperature coefficient always reverses at some high temperature, however, as scattering on phonons necessarily becomes dominant. The considerable dispersion of the experimental resistivity values measured on various samples is also an indication of the finiteness and high dispersion of artificial crystallite sizes and of the impact of their boundaries on electrical conductivity.

 \section{Absorption efficiencies}
 
\begin{figure}
\resizebox{\hsize}{!}{\includegraphics{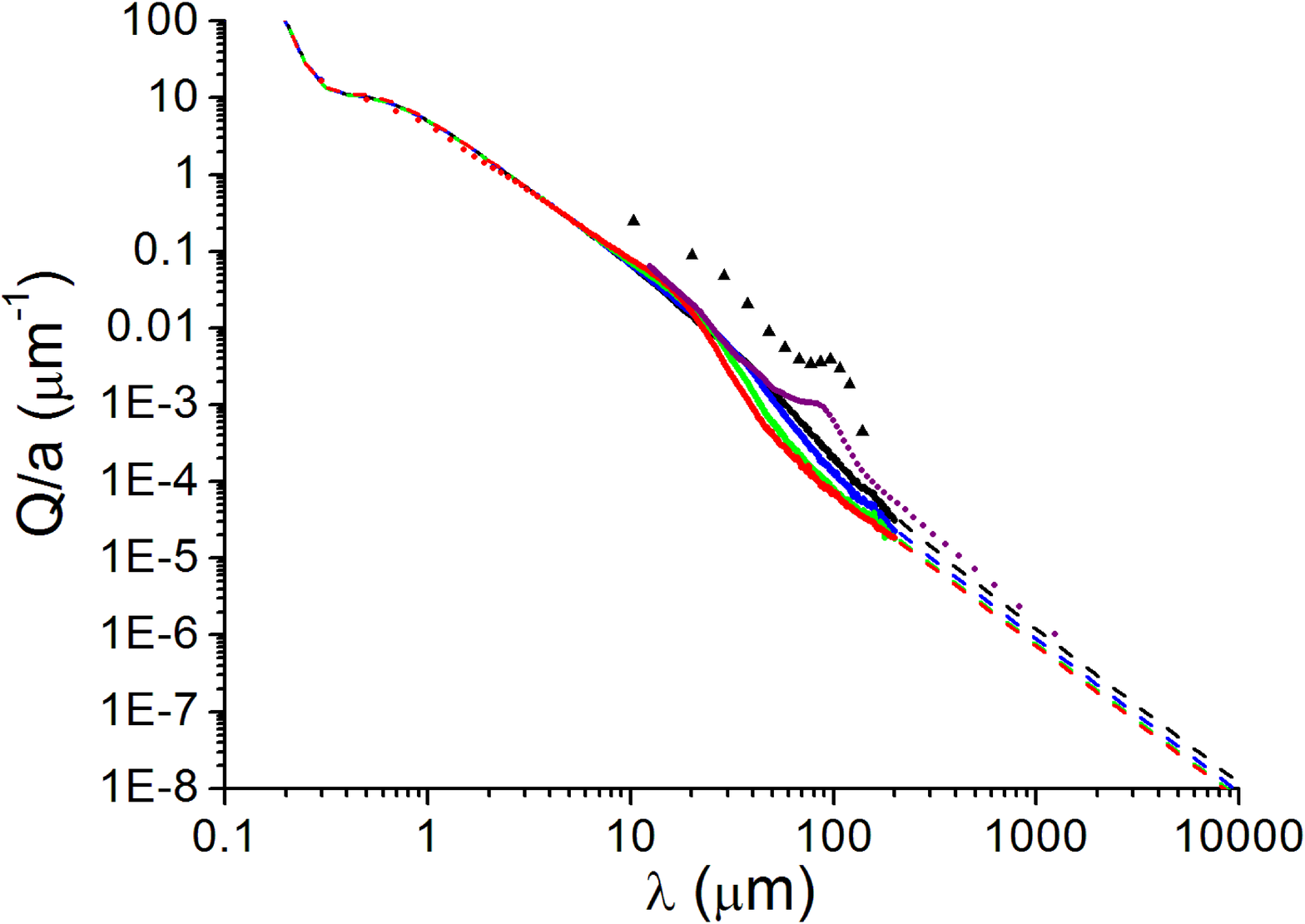}}
\caption[]{The absorption efficiencies computed from the dielectric functions of graphite in the $\vec c \bot \vec E$ orientation, for 10, 100, 200, and 300 K (same conventions as in previous figures). The efficiency decreases as the temperature increases, but only slightly and only beyond $\lambda=10 \mu$m. Also drawn is the curve deduced from the data of Philipp \cite{phi77}, as purple dots, with its characteristic bump around 80 $\mu$m. Triangles represent data deduced from Sato's measurements \cite{sat}, adapted from Koike et al. \cite{koi80}. The particles are assumed to be in the Rayleigh limit.}
\label{Fig:QsA}
\end{figure}

 The model absorption efficiency of grains of radius $a$, small relative to the wavelength $\lambda$, is defined as
 \begin{equation}
\displaystyle
Q/a=\frac{24\pi}{\lambda}\frac{\epsilon_{2}}{(\epsilon_{1}+2)^{2}+\epsilon_{2}^{2}},
 \end{equation}
where $a$ and $\lambda$ are in the same length units. Figure \ref{Fig:QsA} shows  $Q/a$ 
for the same 4 typical temperatures as above. The same quantity directly derived from Kuzmenko's measurements (2008) for each of these temperatures is also shown. Again the agreement between corresponding pairs of curves, in the range where they overlap, is excellent, confirming the consistency of our model.

\begin{figure}
\resizebox{\hsize}{!}{\includegraphics{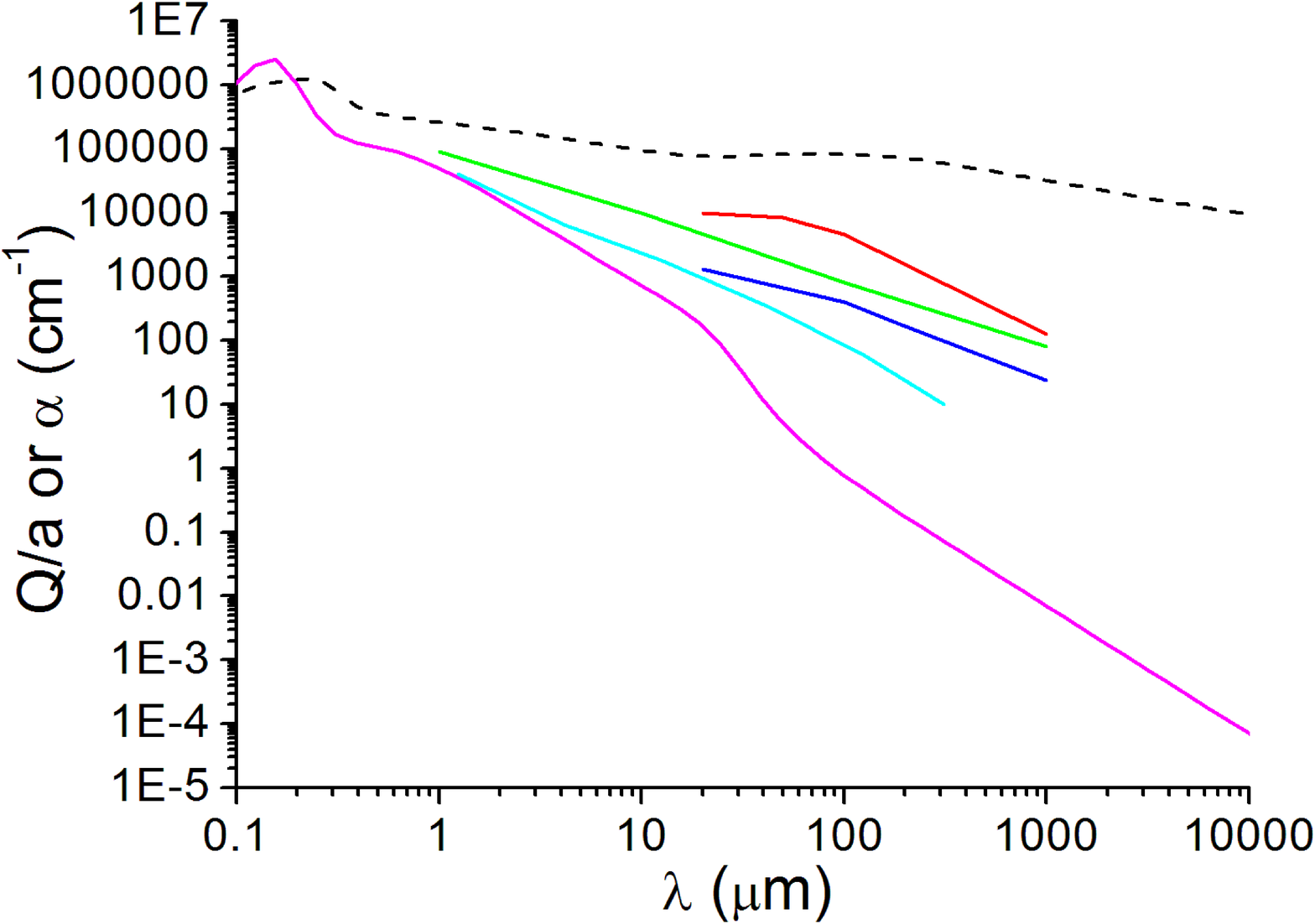}}
\caption[]{Q/a for small oriented spheres of HOPG ($\vec c \bot \vec E$; lower, purple line). Corresponding  absorbance, $\alpha$ (upper, black dashed line). From next to bottom up: $\alpha$ for glassy carbon, adapted from Rouleau and Martin \cite{rou} (cyan line); $\alpha$ for one of the lesser evolved coal, M\'ericourt, adapted from Mennella et al. \cite{men95} (blue line); $\alpha$ for powder graphitic material from electrical discharges in various gases, TH, BE, XY, adapted from Koike et al. \cite{koi80} (green line); $\alpha$ for graphitic powder, adapted from Mennella et al. \cite{men95} (red line). All at room temperature.}               
\label{Fig:alpha}
\end{figure}

 As expected, beyond 10 $\mu$m, $Q/a$ decreases as the temperature rises because the free electrons shield the material from the exciting electric field and more energy is scattered out, so that energy is conserved. For the same reason, the reflectance of the bulk material increases in the same spectral range (Fig. \ref{Fig:R}). By contrast with $Q/a$, the absorbance $\alpha$ of bulk material will be shown below (Fig. \ref{Fig:alpha}) to increase in the same range, as it includes both absorption and reflectance losses. Overall, the grain efficiency scales like $\lambda^{-2}$.

Figure \ref{Fig:QsA} also displays $Q/a$ derived from Philipp's data at room temperature (purple dots). The latter is significantly different than the red dashed and full lines representing the corresponding curves derived from Kuzmenko et al. \cite{kuz} (for reasons stated in the previous section). The figure also includes data displayed in Fig. 1 of Koike et al. \cite{koi80}, which they deduced from Sato \cite{sat}. This is seen to be nearly coincident with Philipp's, to a constant multiplying factor; the difference is discussed by Philipp \cite{phi77}, whose work is based on Sato's FIR reflectance data.
 
 Several measurements were previously performed on essentially graphitic, but disordered, materials in the form of more or less fine grains (see Koike et al. 1980 : TU, BE, XY; Mennella et al. 1995: powder graphite, AC, BE, BS, Mericourt coal). Some of these are displayed in Fig. \ref{Fig:alpha}, for comparison. The figure also includes glassy carbon data derived by Rouleau and Martin \cite{rou}; this material can be imagined in the form of ribbons of graphite (Robertson 1985). Also shown are our model $\alpha$ and $Q/a$ for 300 K. The latter two are seen to bracket all the former.

This variety of optical data on essentially the same material is most likely due to the variety of physical structures in which different preparations deliver the samples, ranging from small, pure, isolated spheres (described by $Q/a$) to bulk pure material (described by $\alpha$). The computations of Rouleau and Martin \cite{rou} convincingly demonstrated how $\alpha$ varies with intermediate structures, like continuous distributions of randomly oriented ellipsoids (CDE), fractal clusters of ellipsoids (FC) and homogeneous porous aggregates. These are accompanied by smaller power indexes, $\beta$, as well as absorbance enhancements relative to isolated spheres, with $\beta$ ranging roughly between 0.6 and 1.25. Figure \ref{Fig:alpha} shows a similar variety of absorbances and power indexes, covering a significant part of the interval between the curves for $\alpha$ and $Q/a$, again for essentially the same material, graphite (at the microscopic level).

This behavior can be understood by considering the large difference between $Q/a$ and $\alpha$, which is small in the UV, and increases steadily with $\lambda$. It is due to the increased shielding of the isolated grain interior from the external field, by the surface charges (see Bohren and Huffman 1983). As the grains grow larger and/or closer to one another, the shielding becomes less and less effective due to the inter-neutralization of adjacent surface charges, resulting in increased absorption $\alpha$. In the limit of the sample of continuous solid material, there is no shielding at all. Now, the data in Fig. \ref{Fig:alpha} were obtained, not on isolated particles, but on pressed pellets of such particles. One cause of the observed behavior of the experimental data may therefore be the more or less dense packing of the powder grains.

Disentangling the dielectric properties of pristine graphite from the effects of size, shape, orientation, agglomeration and aggregation requires inverting the mathematical operations of Rouleau and Martin, which  is obviously a very tricky business. Derivation from reflectance measurements, as in the present work, appears to be an appealing alternative.

\section{Conclusion}

The quantitative results of the present work indicate that small, isolated grains of pure graphite are not suitable to model the long wavelength spectrum of IS emission for a) its emissivity is too low, b) its $\beta$ is too high in absolute value. However, these results should be of help in updating the computed properties of clusters, aggregates and composite grains. 

On the other hand, really (microscopically) amorphous carbons have recently shown a better potential for modeling IR emission beyond 10-20 $\mu$m. Examples are HAC or a-C:H amorphous carbons (Jones et al. 2013) or kerogen/CHONS (Papoular 2014). When associated with amorphous silicate, the latter material ($\beta\sim1.4$; see Papoular 2014) was shown to quantitatively fit, in particular, the DGISM observed by the \emph{Planck} satellite.

 When discussing the relevance of graphite to FIR astrophysics, one must take into account all possible orientations of graphitic dust. Now, the measurement of the dielectric properties of graphite when $\vec E//\vec c$ is much more difficult than when $\vec E\perp\vec c$ as in the present work. Thence the dearth of results on the former orientation, except in DC: see the literature cited in the previous section. The consensus is that, roughly, the ratio of resistivities in the two orientations is about 100. If this ratio is assumed to hold all through the FIR, then the absorption efficiency for $\vec E//\vec c$ is expected to be 100 times higher than for $\vec E\perp\vec c$. This would bring it to about 100 cm$^{-1}$ at $\lambda=1000$ \AA{\ } (see Fig. \ref  {Fig:alpha}), opening up the possibility that it may contribute to the Diffuse Galactic FIR continuum.

The present model of graphite uses only one Lorentzian and 3 Drude functions. It was developed on the basis of new developments in the electronic band theory of graphite. It delivers optical properties of graphite at normal incidence, which comply with the available measurements on HOPG from 10 K to room temperature and up to about $\lambda=200\,\mu$m. The model extends this range into the FIR and provides analytical expressions for all optical properties, from 0.3 to 10000 $\mu$m. In particular, the predicted DC conductivity at room temperature is 1.63$\,10\,^{16}\,$s$^{-1}$, and compatible with the measured value. Also, $Q/a$ in the FIR decreases as temperature increases, has a wavelength power index of $\sim2$ and, at 1000 $\mu$m, is found to be 7$\,10^{-7}$ and 1.4$\,10^{-6}\mu$m$^{-1}$ at 300 and 10 K, respectively. The behavior at room temperature differs considerably from that of the absorbance of powder samples of graphitic materials; a possible reason for this is proposed.

\section{Acknowledgments}
We are deeply grateful to Dr A. Kuzmenko for providing the experimental files, as well as for helpful criticism and advice. Thanks are also due to Dr A. Jones, the reviewer, whose wise comments helped us improve this paper.

\section{Appendix}
The following table completes Table 1 above. Here, the 3 parameters defining the Lorentzian model for the $\pi$ resonance are omitted as they are not changed, assuming they are insensitive to the temperature.

\begin{center}
%%%%%%%%%%%%%%%%%%%%%%%%%%%%%%%%%%%%%%%%%%%%%%%%%%%%%%%%%%%%%%%%%%
\newpage
{\bf Table IA.} Fitting parameters for all measured temperatures\\

\vspace{0.25 cm}
\begin{tabular}{||r||c|c|c|c|c||}
\hline
\hline
T [K] & {
              ${\omega}_p$
            } [eV]             & 
                                      {$\gamma$} [eV]  & 
                                                                   C 
                                                                              & ${{\gamma}_1}$ [eV]  
                                                                                                                    & ${{\gamma}_2}$ [eV]
                                                                                                                                                       \\ 
\hline
\hline
 10 & 0.6179 & 0.0048 & 0.9648 & 0.0048 & 10.0         \\ \hline
 20 & 0.6174 & 0.0048 & 0.9643 & 0.0048 & 10.0         \\ \hline
 30 & 0.6137 & 0.0045 & 0.9600 & 0.0045 & 10.0         \\ \hline
 40 & 0.6158 & 0.0044 & 0.9577 & 0.0046 & 10.0         \\ \hline
 50 & 0.6486 & 0.0047 & 0.9572 & 0.0079 & 10.0         \\ \hline
 60 & 0.6678 & 0.0048 & 0.9579 & 0.0105 & 10.0         \\ \hline
 70 & 0.6903 & 0.0051 & 0.9603 & 0.0129 & 10.0         \\ \hline
 80 & 0.7039 & 0.0050 & 0.9629 & 0.0146 & 10.0         \\ \hline
 90 & 0.7239 & 0.0051 & 0.9669 & 0.0169 & 10.0         \\ \hline
100 & 0.7381 & 0.0052 & 0.9693 & 0.0193 & 10.0         \\ \hline
110 & 0.7563 & 0.0052 & 0.9716 & 0.0217 & 10.0         \\ \hline
120 & 0.7694 & 0.0054 & 0.9759 & 0.0239 & 10.0         \\ \hline
130 & 0.7810 & 0.0056 & 0.9798 & 0.0263 & 10.0         \\ \hline
140 & 0.7946 & 0.0057 & 0.9843 & 0.0288 & 10.0         \\ \hline
150 & 0.8067 & 0.0057 & 0.9871 & 0.0311 & 10.0         \\ \hline
160 & 0.8186 & 0.0058 & 0.9911 & 0.0333 & 10.0         \\ \hline
170 & 0.8295 & 0.0058 & 0.9964 & 0.0357 & 10.0         \\ \hline
180 & 0.8402 & 0.0058 & 1.0009 & 0.0378 & 10.0         \\ \hline
190 & 0.8490 & 0.0060 & 1.0039 & 0.0403 & 10.0         \\ \hline
200 & 0.8591 & 0.0060 & 1.0079 & 0.0427 & 10.0         \\ \hline
210 & 0.8680 & 0.0060 & 1.0110 & 0.0446 & 10.0         \\ \hline
220 & 0.8769 & 0.0060 & 1.0149 & 0.0472 & 10.0         \\ \hline
230 & 0.8848 & 0.0059 & 1.0174 & 0.0489 & 10.0         \\ \hline
240 & 0.8938 & 0.0061 & 1.0241 & 0.0514 & 10.0         \\ \hline
250 & 0.9008 & 0.0062 & 1.0262 & 0.0535 & 10.0         \\ \hline
260 & 0.9122 & 0.0063 & 1.0358 & 0.0561 & 10.0         \\ \hline
270 & 0.9211 & 0.0065 & 1.0420 & 0.0585 & 10.0         \\ \hline
280 & 0.9270 & 0.0065 & 1.0466 & 0.0610 & 10.0         \\ \hline
290 & 0.9327 & 0.0067 & 1.0521 & 0.0635 & 10.0         \\ \hline
300 & 0.9441 & 0.0068 & 1.0364 & 0.0627 & 10.0         \\ 
\hline
\hline
\end{tabular}
%%%%%%%%%%%%%%%%%%%%%%%%%%%%%%%%%%%%%%%%%%%%%%%%%%%%%%%%%%%%%%%%%%
\end{center}

\end{document}